# Deeply-trapped molecules in self-nanostructured gas-phase material

**M. Alharbi[1,2], A. Husakou[1,3], B. Debord[1], F. Gerome[1] and F. Benabid[1,2]**

[1]GPPMM group, XLIM Research Institute, CNRS UMR 7252, University of Limoges, France
[2]Physics department, University of Bath, Bath, UK
[3]Max Born Institute of Nonlinear Optics and Short Pulse Spectroscopy, Berlin, Germany
f.benabid@xlim.fr

**Since the advent of atom laser-cooling, trapping or cooling natural molecules has been a long standing and challenging goal. Here, we demonstrate a method for laser-trapping molecules that is radically novel in its configuration, in its underlined physical dynamics and in its outcomes. It is based on self-optically spatially-nanostructured high pressure molecular hydrogen confined in hollow-core photonic-crystal-fibre. An accelerating molecular-lattice is formed by a periodic potential associated with Raman saturation except for a 1-dimentional array of nanometer wide and strongly-localizing sections. In these sections, molecules with a speed of as large as 1800 m/s are trapped, and stimulated Raman scattering in the Lamb-Dicke regime occurs to generate high power forward and backward-Stokes continuous-wave laser with sideband-resolved sub-Doppler emission spectrum. The spectrum exhibits a central line with a sub-recoil linewidth of as low as 14 kHz, more than 5 orders-of-magnitude narrower than in conventional Raman scattering, and sidebands comprising Mollow triplet, molecular motional-sidebands and four-wave-mixing.**

Engineering the interaction between optical fields and gas-phase matter is at the core of many key fields in modern physics, such as high resolution spectroscopy, quantum optics and laser science and technology. However, because of atoms or molecules random thermal motion and delocalization, their spectral signature is Doppler broadened, hence obscuring their quantum dynamics and destroying their coherence. As a result, probing and controlling these atoms/molecules coherently and at the single entity level has always been very challenging. Among the numerous ingenuous laser-matter interaction schemes to reach the sub-Doppler resolution and penetrate the single atom/molecule dynamics, we count that of laser cooling techniques[1–3], which led to atomic Bose-Einstein condensate (BEC) [4,5]. The latter enabled the observation with unparalleled details of several sub-Doppler quantum phenomena such as spectral signatures from absorption and emission with sub-recoil linewidth [6] and resolved atomic translational motions, Rabi splitting or Mollow triplet, and collective effects induced by dipole-dipole interaction to mention a few [1–3]. The observation of these phenomena comes, however, at the expense of highly complex experimental systems, and it is often the case that a sophisticated set-up is dedicated to the sole purpose of observing a single effect. Furthermore, this rich dynamics has so far been limited to atoms, since the extension to natural molecules proved to be extremely challenging to implement due to their much more richer internal structure[7]. Given the potential of trapped and cooled molecules in shedding new light on fundamental phenomena such as long-range dipole interactions, parity violation, and chemical process dynamics to mention a few, a considerable endeavour is currently being undertaken to trap and cool molecules with the advent of several techniques such as electrostatic [8] or optical [9] Stark decelerator, and buffer gas cooling [10].

Here, we report on a radically new means of trapping extremely high pressure Raman-active molecules in a self-assembled optical lattice whose depth of ~55 THz allows the staggering velocity capture of ~1800 m/s. Here, the molecules are hydrogen in photonic bandgap guiding hollow-core photonic crystal fibre (HC-PCF)[11] excited with a powerful continuous-wave laser. The molecular trapping mechanism contrasts with previous techniques by the fact that the molecules create their own optical lattice by exciting forward and backward Stokes waves to form a standing wave along the fiber. Concomitantly, in the nodes of this standing wave the molecules are tightly trapped in the low-energy ground quantum-state, while the remaining molecules are strongly saturated and in the high energy state. The outcome is an uncommon self-optically nano-structured gas-phase material (SONS-GPM) that consists of several centimetre long 1-dimentional array of nano-localized Raman-active molecules that scatters light in the Lamb-Dicke regime. As a result of this dynamics, we observed the remarkable generation of continuous-wave Stokes laser that combines a power level of up to 50 W and a sub-recoil linewidth of down to 14 kHz. The Stokes emission spectrum has a sub-Doppler structure that was thought to solely occur in ultracold atoms. It exhibits resolved sidebands due to Rabi splitting, trapped molecule translational motion, and shows strong optical nonlinearity via strong four-wave-mixing (FWM) inter-sidebands. A further striking property of this molecule-laser interaction is that the formed potential is an accelerating optical potential that imparts a molecular drift, which is dramatically observable with the naked eye via IR viewer. Below we organize the article by starting with a description of the experimental arrangement, we then give a theoretical background explaining the unusual results to finish with a detailed experimentally-backed description of SONS-GPM microscopic and macroscopic dynamics.

**Experimental configuration for Stokes generation**

Fig. 1A schematically shows the experimental set-up (see Methods, Appendix A for more detail). A 20 m long home-made photonic bandgap (PBG) guiding HC-PCF (top of Fig. 1B), filled with molecular hydrogen[13] at uniform pressure in the range of 20-50 bar, is pumped with a randomly polarized 1061 nm wavelength Yb-fibre CW laser with a linewidth of only ~400 kHz. In such a system, stimulated Raman scattering (SRS) occurs whereby the incoming input laser photons transform into lower-energy (longer-wavelength) Stokes photons by transferring their energy difference to molecular rotational and/or vibrational transitions. The narrow transmission window of the PBG HC-PCF (red curves of Fig. 1C) enables the excitation of only the rotational transitions $S_{00}(1)$ of $H_2$ with a frequency-shift of $\omega_R \sim 2\pi \times 17.8\,\mathrm{THz}$[14], and hence set the molecule in a well-defined quantum Raman state.

The blue curves in Fig. 1C show the typical recorded optical spectra of the transmitted forward and backward spectrum beams for pump power of 29 W (i.e. ~17.5 W of coupled power) and hydrogen pressure of 20 bar. The forward spectrum shows two spectral lines; one at 1061 nm from the pump laser (P), and a second at 1131.4 nm from forward 1$^{st}$ order Stokes (FS). The backward spectrum is composed mainly of the backward Stokes (BS) line, with residual light from the pump which is back reflected off the fibre input end. Fig. 1B also shows the typical fibre output beam profile at the pump (middle of Fig. 1B) and the Stokes wavelength (bottom of Fig. 1B). Fig. 1D shows RF spectra of typical linewidth traces over $\pm 2$ MHz spectral span of P, FS, and BS. The linewidth of the fibre-transmitted pump was measured to be ~ 400 kHz, and is consistent with the manufacturer specifications. This linewidth value and spectral-profile remain unchanged when both the fibre-coupled power and the gas pressure are varied (Methods, Appendix B).

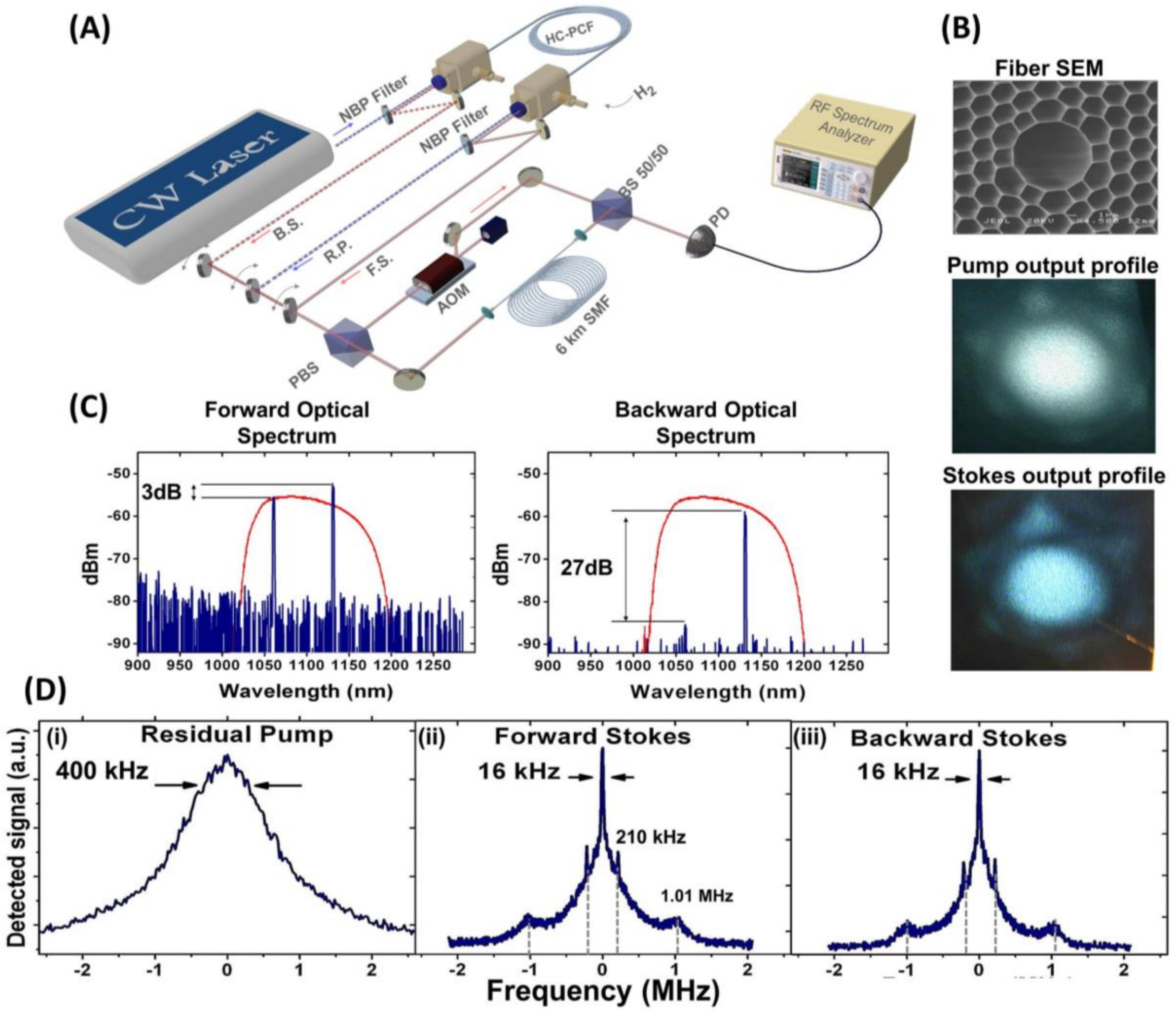

**Fig. 1 Stokes generation and linewidth measurement** (a) Schematic of experimental setup consisting of Raman generation through PBG HC-PCF and self-heterodyne linewidth measurement technique. NBP Filter: narrow bandpass filter; BS: Backward Stokes; FS: Forward Stokes; RP: residual pump, AOM: acousto-optic modulator; PBS: polarization beam splitter; BS: beam splitter; SMF: single mode fibre; PD: photodetector. (B) Fibre scanning electron micrograph (top), laser beam profile of the transmitted pump ( middle) and the generated forward Stokes (bottom). (C) Optical spectra of the fibre transmission, stimulated Raman scattering spectra generated in the forward direction (lhs), and backward direction (rhs). Self-heterodyne RF spectra showing the linewidth of the transmitted residual pump (i), forward Stokes (ii) and backward Stokes (iii).

The FS and BS linewidth traces, however, not only depend on the coupled pump power and the gas pressure (see below), but dramatically contrast with what one would expect from the conventional physical picture of SRS process, which predicts a pressure broadening Stokes linewidth of around 2 GHz for a gas pressure of 30 bar [15]. The present emitted Stokes lines stand out with a linewidth that is over 5 orders of magnitude narrower than the pressure broadened line. Here, the measured linewidth of $\Delta\nu$ =16 kHz for both FS and BS are well below the Doppler-limit linewidth of 153 MHz and 300 MHz for $H_2$ rotational Raman transition in forward and background configurations respectively[16]. Most remarkably, the measured linewidth is below the hydrogen molecule recoil frequency $\nu_{recoil} = \pi\hbar/(m{\lambda_S}^2)$ of ~78 kHz. Furthermore, FS and BS linewidth spectra show two pairs of sidebands; one extremely narrow with a frequency shift from the central-peak of $210 \pm 15$ kHz, and a wider pair with 1.01 MHz frequency-shift. These spectral characteristics are usually signatures of strongly-trapped atoms in the Lamb-Dicke regime [17]. In our case, this Lamb-Dicke regime results from an original dynamics, which we outline below.

**Theoretical background**

The present laser-gas interaction is modelled (see Methods Appendix C for a detailed account of the model and the numerical simulations) by treating the Raman medium as a two-level system [18] $\{|1\rangle,|2\rangle\}$. Here, $|1\rangle$ and $|2\rangle$ represent the ground and excited states of the molecular rotation respectively (Fig. 2D). The coupling between the two states is defined by the two-photon Rabi frequency $\Omega_{12} = \Omega_{21}^* = (1/2)d_S E_p E_S^*$. Here, $E_p$ and $E_S$ are the pump and Stokes fields respectively, and $d_S$ is the coupling coefficients between the two states[19]. In our case, the Stokes fields are amplified from the quantum noise, and is generated in both forward and backward direction relative to that of the pump. Consequently, $\Omega_{12}$ is spatially modulated following the expression:

$$|\Omega_{12}|^2 = {\Omega_{12}^{(0)}}^2 \left(1 + r_S^2 + 2r_S\cos(2\beta_S z)\right)/2. \quad (1)$$

Here, $\Omega_{12}^{(0)} = \left|d_S \bar{E}_p\ \bar{E}_S^{(f)}/\sqrt{2}\right|$, $\bar{E}_p = E_p e^{i\beta_p z}$, $\bar{E}_S^{(f)} = E_{FS} e^{i\beta_S z}$, $\bar{E}_S^{(b)} = E_{BS} e^{-i\beta_S z}$, and $E_S = E_S^{(f)} + E_S^{(b)}$. $\beta_{p(S)}$ is the propagation constant for the pump (Stokes). $r_S$, defined by $\left|E_S^{(b)}\right| = r_S \left|E_S^{(f)}\right|$, is the local ratio between the forward and backward components of the pump or Stokes field. The superscript $(f)$ and $(b)$ stand for forward and backward. We ignored the backward pump as it is too small to have meaningful effect. In turn, such a $z$-dependent periodic structure is exhibited by the population difference, $D$, Raman gain, $g_R$ and the expectation-value $\langle U_{tot}\rangle$ of the potential as follows:

$$\begin{cases} D = \rho_{22} - \rho_{11} = -1/\left[1 + 4\,|\Omega_{12}|^2/(\Gamma_{12}\gamma_{12})\right], \\ g_R = g_R^{(0)} D\left[{\gamma_{12}}^2 + (\Omega_{11} - \Omega_{22})^2\right]^{-1}, \\ \langle \hbar^{-1} U_{tot}\rangle = -D|\Omega_{12}|^2 \left(\frac{\Omega_{11}-\Omega_{22}}{\gamma_{12}^2+(\Omega_{11}-\Omega_{22})^2}\right) + \left(\frac{1-D}{2}\Omega_{11} + \frac{1+D}{2}(\Omega_{22} + \omega_R)\right). \end{cases} \quad (2)$$

Here, $\rho_{11}$ and $\rho_{22}$ are the diagonal elements of our two-level Raman system density matrix. $\gamma_{12} \sim 2\pi \times (p_g/20) \times 1.14\,GHz$ [20], and $\Gamma_{12} \sim 2\pi \times (p_g/20) \times 20kHz$ [21] are the pressure-dependent for thermal molecules Raman dephasing-rate and population decay-rate respectively. $\Omega_{11} = (1/2)(a_S E_S E_S^* + a_p E_p E_p^*)$, $\Omega_{22} = (1/2)(b_S E_S E_S^* + b_p E_p E_p^*)$ are the Stark shift of the ground and the excited state. Quantities $a_{p(S)}, b_{p(S)}$ are the coupling coefficients related to the dipole moments between the different states involved (including the electronic upper states) [19]. Here, $g_R^{(0)} = -2\hbar N d_S^2 \omega_S (n_S n_p \gamma_{12} c^2 \varepsilon_0^2)^{-1}$, with $N, n_S, n_p, c$ and $\varepsilon_0$ are the molecular density, the Stokes and pump refractive indices, the vacuum light speed and the vacuum permittivity respectively. In addition to this spatial periodicity, two further salient features give the above physical quantities a unique periodic spatial distribution. First, in our experimental conditions the saturation condition ($i.e.\ \Omega_{12}^{(0)} \geq \Omega_{12}^{(sat)} = (1/2)\sqrt{\gamma_{12}\Gamma_{12}}$) is reached with an ultra-low power-level. This results from the strong transverse confinement of the fibre guided-mode. For example with a pump power of 0.5 W (*i.e.* intensity of 3.3 MW/cm$^2$), and Stokes power of 0.25 W, the corresponding two-photon Rabi frequency reaches a value of $2\pi \times 10MHz$, which is over four times larger than $\Omega_{12}^{(sat)} \sim 2\pi \times 2.4\ MHz$ for a a gas pressure of 20 bar. Second, the dominant contribution to the Hamilton, which is provided by the term $D\omega_R/2$ leads to a large potential depth between ground-state molecules (*i.e.* $D = -1$) and those that are Raman-saturated

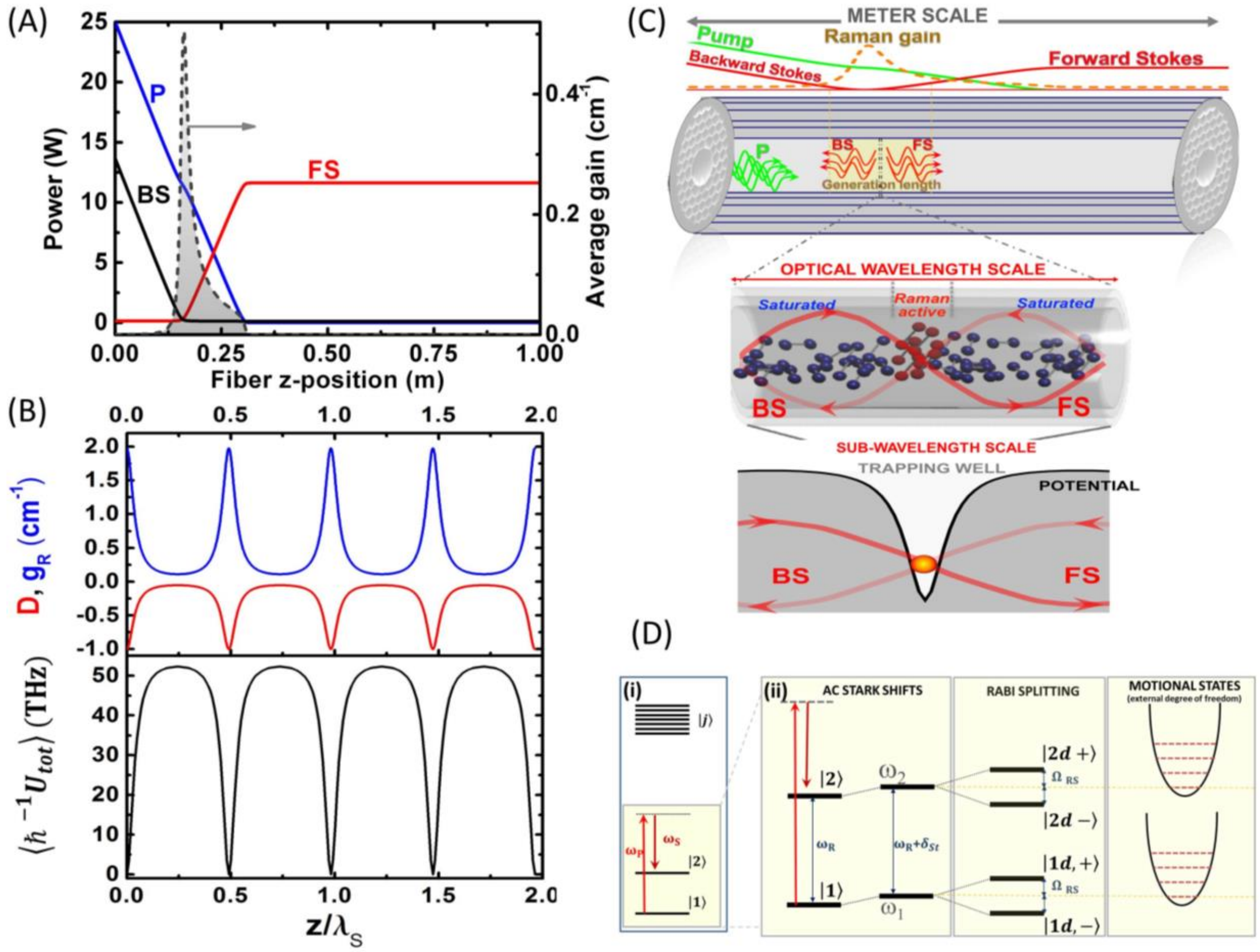

**Fig. 2 Self-optically nano-structured Raman-active molecules.** (A) Macroscopic power z-distribution along the fibre of the pump (blue curve), FS (red curve) and BS (black curve). The grey dashed and filled curve is the average Raman gain (in unit of cm$^{-1}$). (B) Microscopic z-distribution (normalized to the Stokes wavelength) of the Raman gain (blue curve), the normalized population difference (red curve), and the expectation value of the potential (black curve). (C) Schematics of SONS-GPM dynamics and scale from the meter-scale of the hydrogen-filled fibre (top),to a single Stokes-wavelength scale (middle) to then nanometer-scale (bottom). The red spheres represent the Raman active molecules and the blue ones the Raman saturated molecules. The pump is represented by a green curve and letter P. FS and BS by red curves. (D) Schematics of the energy level of the molecules inside the SONS-GPM both for the internal degree of freedom (here hydrogen molecule rotation) (i), their Stark shift and Rabi splitting along with the external degree of freedom (translational motion of the molecules) (ii).

molecules (*i.e.* $D = 0$).

**Molecular macroscopic and microscopic structure**

Fig. 2 macroscopically and microscopically captures this self-induced structuring, via numerical calculations for a representative case of $P_{in} = 25$ W and $p_g = 20$ bar (Fig. 2A and 2B), and schematically (Fig. 2C and 2D). Fig. 2A shows the numerically calculated power distribution along the fibre length of the pump (P), forward Stokes (FS), backward Stokes (BS) and the Raman average gain $\langle g_R \rangle$ respectively. $\langle g_R \rangle$ is expressed in units of $cm^{-1}$ with an average value of $\langle g_R \rangle_{avg} \sim 0.2\ cm^{-1}$ and a maximum of $\langle g_R \rangle_{max} \sim 0.5\ cm^{-1}$. At this macroscopic scale, the evolution of P, FS, BS and $\langle g_R \rangle$ with the propagation distance is typical to the conventional picture of SRS process [22]. The pump-power is gradually depleted, and within the Raman generation length $l_R \sim \langle g_R \rangle_{avg}^{-1}$, corresponding to the length of the filled dashed-curve in Fig 2A, a quick rise of FS and BS takes place and a strong Raman gain peak develops. Outside this region, the Raman gain drops to almost zero. This macroscopic dynamics is schematically illustrated in the top of Fig. 2C. At the microscopic scale, however, the results contrast with prior work by having $D, g_R$ and $\langle \hbar^{-1} U_{tot} \rangle$ exhibiting a periodic and spatially nanostructured z-distribution that extends over the whole Raman generation length. The three quantities exhibit a spatial modulation with a period of $\lambda_S/2$ and comprising two distinctive alternating regions (see Fig. 2B). One is extremely narrow and corresponds to local high gain Raman-active region, and a second one, relatively wide, corresponding to a vanishing gain Raman-saturated region. The latter corresponds to a strong SRS suppression with $D$ and subsequently $g_R$ decreases to almost zero. On the other hand, the Raman-active region is characterized by a sub-wavelength-wide region whose spatial-FWHM can be written as $\delta z = \left(\lambda_S/\pi\sqrt{2}\right)\left(\Omega_{12}^{(sat)}/\Omega_{12}^{(0)}\right)$ for the case of $r_S = 1$. At the centre of the Raman-active region the local Raman gain reaches its maximum and the population difference remains dominantly in the ground-state. Remarkably, the RA region is also the site of an extremely deep trapping-well with a depth along the z-direction, $\delta\langle \hbar^{-1} U_{tot} \rangle \sim \omega_R/2$ of more than 55 THz corresponding thus to the staggering figures of velocity-capture of 1800 m/s and effective temperature of ~425 K. This 55 THz potential-depth is several orders of magnitude larger than the few MHz potential depth, which is typically used in optical atom traps. It also leads to the approximate fourfold increase of the molecules density in the nano-sections as estimated by Boltzmann distribution, partly explaining the strong quantum conversion to the Stokes (Methods, appendix I). This SONS-GPM is schematically illustrated in the bottom of Fig. 2C where the scattering molecules (red-colored) are "sandwiched" between saturated molecules (blue-colored) within the nano-wide section and are kept there with the trapping potential forces.

Furthermore, Fig. 2D illustrates the expected changes to energy levels within this regime of sub-wavelength localization and strong driving fields. Each dressed-state of the molecular rotation should exhibit quantized translational motion-states represented in Fig. 2D(ii). In our case, the potential 3D profile for the representative case of $P_{in} = 25$ W and $p_g = 20$ bar [Fig. 3A(i)] indicates two motional states. First, a longitudinal trapping provided by well-depth $\delta\langle \hbar^{-1} U_{tot} \rangle \sim \omega_R/2$ of more than 55 THz [Fig. 3A(ii)] and corresponding to a longitudinal motion with a frequency $\nu_{long} = 2\left(\Omega_{12}^{(0)}/\Omega_{12}^{(sat)}\right)\sqrt{\nu_{recoil}\langle \hbar^{-1} U_{max,l} \rangle / \pi}$ of ~7GHz. Second, the potential provides transverse trapping with a well-depth of ~200 MHz (Fig. 3A(iii)), corresponding to transverse oscillations at a frequency $\nu_{trans} =$ $(2.405/2\pi r)\sqrt{\langle U_{max,t}\rangle/m}$ of ~300 kHz. Here, $U_{max,l}$ and $U_{max,t}$ are the potential local-maxima along z-direction and transverse direction respectively.

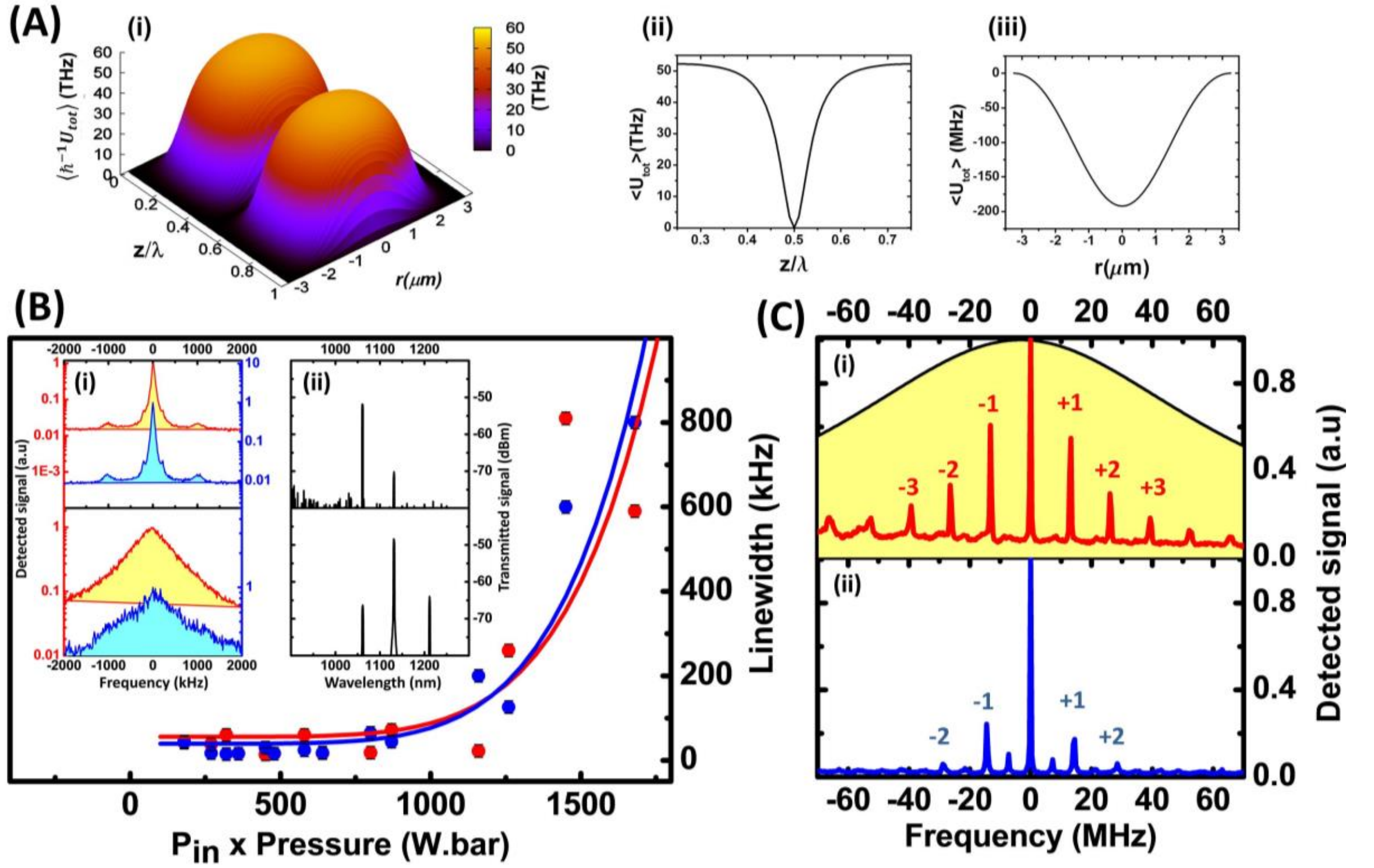

**Fig.3 Measured Stokes-linewidth spectral-structure.** (A) Microscopic 3D profile of the Hamiltonian (i). Hamiltonian longitudinal profile (ii) and transverse profile (iii) in the vicinity of an anti-node (i.e. at $z = n\,\lambda_S/2$, with $n$ being an integer). (B) FS (red color) and BS (blue color) linewidth evolution with $P_{in} \times p_g$. The points are experimental data and the solid lines are fitted curves to the experimental points. Inset: Linewidth spectral traces (i) and transmitted optical spectra (ii) for $P_{in} = 9\ W$, and $p_g = 20\ bar$ (top), and for $P_{in} = 29W$, and $p_g = 50\ bar$ (bottom). (C) FS (red curve), and BS (blue curve) spectrum over 150 MHz span. The BS Doppler-limited linewidth (yellow filled-curve) is shown for comparison.

## Line narrowing and sideband generation dynamics

This picture of deeply trapped scattering molecules, which is further corroborated below, explains the observed combination of a strong conversion to the Stokes (Fig. 1C), the narrow linewidth (Fig. 1D) and the resolved sidebands.

i) ***Line narrowing:*** Indeed, with such potential, the expected Lamb-Dicke [17] narrowing of its Lorentzian-shape spectral signature scales with the ratio $\sqrt{\delta\langle\hbar^{-1}U_{tot}\rangle/2\pi\nu_{recoil}}$ [23] of $1.02\text{x}10^4$, yielding a linewidth of 15 kHz, which is in very good quantitative agreement with the measured FS linewidth narrowing relative to the Doppler broadening (Fig. 2D). The strong Lamb-Dicke line narrowing is observed for different pump laser powers and gas pressures and are shown in Fig. 3B. The ranges investigated were between $P_{in} = 9$ W (*i.e.* ~6 W of coupled power) and $P_{in} = 42$ W for the laser power, and $p_g = 20$ bar to $p_g =$ 50 bar for the gas pressure (see Methods, appendices D and E for larger data set). Fig. 3B shows the dependence with the gas pressure and pump input-power of FS and BS linewidth, and hence that of the SONS-GPM dynamics. Within our explored input power and gas pressure ranges, we found that for $P_{in} \times p_g \lesssim 1000\, W\, bar$, the linewidth varies little and reaches a minimum of 14 kHz and a maximum of 67 kHz. The small variation in linewidth in our explored range of pump power and gas pressure is explained by the fact that the potential depth $\delta\langle \hbar^{-1} U_{tot}\rangle \sim \omega_R/2$ is dominated by the Raman transition frequency, with the Stark and Rabi frequencies being several orders of magnitude smaller. Above $1000\, W\, bar$, the appearance of the second-order Stokes, as visible in the transmitted optical spectrum (see bottom of Fig. 3B(ii)) alters the dynamics and leads to broader Stokes linewidth exhibiting a "triangular" profile with the sidebands no longer resolved (bottom of Fig. 3B(i)). It is noteworthy however, that one can scale the fibre length and pressure so as to operate in the above-mentioned Lamb-Dicke regime with a larger input power level and close to quantum conversion to the Stokes (Methods, Appendix F).

ii) ***Transverse motion sidebands:*** Furthermore, in addition to the Lamb-Dicke Lorentzian central-peak, the 4 MHz span RF spectra exhibit, as mentioned above, two pairs of sidebands located respectively at roughly $\pm 210$ kHz, and $\pm 1$ MHz from the central peak [see top of Fig. 3B(i)]. The $\pm 210$ kHz sidebands are consistent with the transverse motional frequency $\nu_{trans}$ which was found to be in the range of 200-400 kHz range for pump power and gas pressure.

iii) ***Rabi splitting:*** Fig. 3C shows FS and BS emission spectral-structure over a larger span of 150 MHz for the case of $P_{in} = 16\, W$, and $p_g = 50\, bar$. In addition to the sub-recoil narrow central-peak, the spectra exhibit several strong equally spaced sidebands from the central peak (labelled $\pm 1, \pm 2, \ldots$ in Fig. 3C). The corresponding FS Doppler-limited linewidth is shown with yellow filled curve to put into evidence the sub-Doppler regime of the generated spectrum [Fig 3C(i)]. We recall that under a strong pump, the unperturbed energy levels of the conventional two-level system [Fig. 2D(i)] are AC-Stark shifted and Rabi split [Fig. 2D(ii)]. The system is now better described by a 4-level system dressed states[1] $\{|1d,-\rangle, 1d,+\rangle, |2d,-\rangle, 2d,+\rangle\}$ [Fig 2D(ii)], and the scattering spectrum would exhibit Mellow triplet spaced by the two-photon Rabi frequency. The measured FS sideband spacing of 13.2 MHz is in good agreement with the $2\pi \times 10.2\, MHz$ of the numerically calculated effective two photon Rabi frequency $\Omega_{RS}$ at the nano-sections (see Methods Appendix G). The agreement between theory and experiment was also found in the fact that $\Omega_{RS}$ varies little with $P_{in}$. Consequently, the first two lines (*i.e.* labelled $\pm 1$) form with the central line the Mollow triplet. The higher-order lines (labelled $\pm 2, \pm 3 \ldots$) are the results of inter-sidebands FWM that occur via optical nonlinearity [24]. Additionally, the spectrum exhibits weaker sub-harmonic sidebands with half the spacing-frequency, which likely results from quantum interference between the different quantum-paths between the dressed-states [25].

iv) ***Longitudinal-motion sidebands***: Finally, for the present potential depth magnitudes, we have typically $\nu_{long}$ to be in the range of 2-10 GHz and $\nu_{trans}$ in the range of 200-400 kHz. The transverse frequency $\nu_{trans}$ range is in good agreement with the experimentally observed spikes in Stokes linewidth spectrum (Fig. 1D). The longitudinal motional frequency was outside our detection range. However, the examination of the spatial profile of the FS diffracted beam confirms the existence of three strong spectral lines spaced by roughly 5-10$\pm$5 GHz. Follow-up work is underway to measure them with better resolution, but also to explore whether this configuration leads to resolved sideband molecular cooling [26], and hence provide further insight into the dynamics of the molecules within the nano-sections.

## Lattice in motion

Similarly, the BS spectrum (Fig 3C(ii)) exhibits the same corresponding Mollow sidebands and their higher-order lines. However, the spacing between the BS Rabi sidebands was found to be 14.4 MHz, which is 1.2 MHz higher than for FS. This difference results from a Doppler shift between FS and BS, which is due to drift velocity of the SONS-GPM along $z$-direction of $v_z^{(drift)} = \Delta\nu_D \lambda_S/2$ of 0.7 m/s. The dynamics behind the drift is due to the fact that the SONS-GPM opto-molecular lattice is an accelerating lattice (*i.e.* it has Wannier-Stark ladder-structure [12]) and whose macroscopic dynamics are remarkably observable with the "naked eye" as it is shown below. Also, this frequency shift between FS and BS sidebands and their coexistence in the gain region leads to the FWM with the central peak, resulting in the sideband at roughly 1.1 MHz, which is indeed visible in Fig. 3B(i).

The origin and the description of the molecular lattice acceleration are depicted by Figs. 4A-C. Figures 4A presents the z-distribution over the first 35 cm long fibre section of the macroscopic average potential $\langle \hbar^{-1} U_{tot} \rangle$. By virtue of the optical force $F_z = -\nabla_z \langle U_{tot} \rangle$ the molecules would undergo acceleration in the fiber sections labeled FS1 and FS3, and an oscillatory regime in the FS2 region. To have an insight on the molecular dynamics it is instructive to shortly consider their transport under ballistic regime (i.e. the diffusion and viscosity are ignored). The resulting molecules' trajectories diagram is presented in the phase-space at the macroscopic level in Fig. 4B, and at the microscopic level in Fig. 4C (Methods, appendix H). One can clearly see the total drift of the molecules towards the fiber end due to the potential gradient which reaches values up to $\nabla_z \langle \hbar^{-1} U_{tot} \rangle$ =-3070 THz/m, corresponding to ~1 million times the gravitational acceleration on Earth. In addition, in FS2 region the motion develops into a complex oscillatory regime that exhibits two spatial orders: microscopic order set by the extremely narrow (black sections in Fig. 4B ) 55-THz-deep periodic potential (velocity capture of ~1800 m/s) and the macroscopic order set by a trapping potential-well with a velocity capture of ~800 m/s and extending over the FS2 region illustrated by the separatrix contour (yellow closed curve in Fig. 4B).

As a result, the molecules in the fiber experience a global drift towards the fiber output end. This overall flow velocity was numerically calculated by introducing the laminar viscous flow[27] and $H_2$ viscosity value to be ~ 0.3 m/s. This is consistent with the one deduced from the Doppler shift between FS and BS for the molecules inside the FS2 section. A further experimental corroboration of this flow is demonstrated in a dramatically visible fashion by video recording the side-scattered light from a 4 m long HC-PCF section coiled in a spiral for the case of $P_{in} = 25\,W$, and $p_g = 20\,bar$, and situated at z ~ 2 m away from the fibre input (see Methods Appendix J for full video). The video clearly shows the motion of scattering nanoparticles from the fibre gas-cell that are dragged along the hydrogen flow, and which are indicated by yellow-circles in Fig. 4D, which shows a few selected video frames. We observe that the scattering particles velocity is controllable by the input power, as seen in Fig. 4E showing the scatterer position versus time for different input powers. The results indicate that for each input power the scatterer moves with a constant drift velocity in this section of the fibre, in agreement with theory. Fig. 4F shows that if input power is increased from 15 W to 30 W, the measured drift velocity increases from ~1.5 m/s to ~4 m/s respectively. This range is in reasonable agreement with the one deduced from FS-BS Doppler shift. For $P_{in} > 30\,W$, the velocity decreases because of the presence of the second order Stokes. Finally, it is worth noting that the tight-binding limit ($\delta\,\langle\hbar^{-1}U_{tot}\rangle \gg 2\pi\nu_{recoil}$) [12] used above for molecular dynamics within the nano-sections still holds because the macroscopic potential induced $F_z = -\nabla_z\langle U_{tot}\rangle$ remains much smaller than the critical force $F_c = (\pi/32)\left(\langle\hbar^{-1}U_{max,l}\rangle/2\pi\,\nu_{recoil}\right)\left(\langle U_{max,l}\rangle/\lambda_S\right)$ for Landau-Zener tunneling between nano-sections [12].

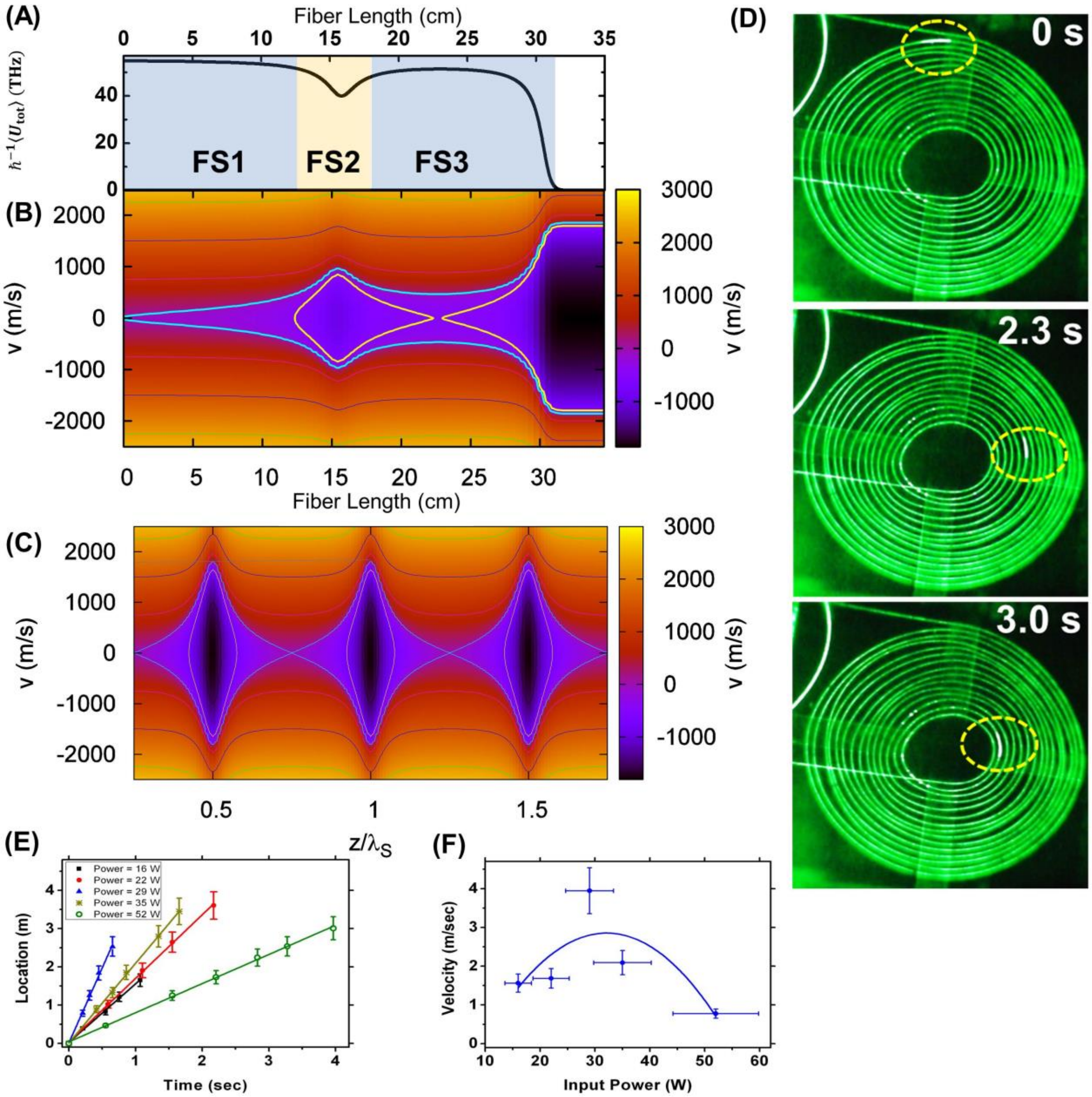

**Fig. 4 Molecular acceleration and macroscopic dynamics.** Macroscopic z-distribution over the first 35 cm HC-PCF of the Hamiltonian (A) and the phase-space diagram (B). FS1, FS2 and FS3 represent the fibre sections where molecules experience different acceleration regimes. (C) Zoomed-in phase-space diagram over $\sim 2\lambda_S$ wide section from FS2. (D) Selected frames from the video (see Methods Appendix J) showing the moving scatter (dashed-circled sections) along the spirally-set fibre. The frame corresponding time sequence is shown on the right top. (E) Scatter's location evolution with time for different input pump powers. (F) Scatter's velocity with input pump power.

## Conclusion

The present configuration of trapping molecules and its intriguingly rich results will undoubtedly provoke thoughts and ideas in further exploring theoretically and experimentally the underlined physics with potential impact on quantum molecular physics comparable to that of laser atom cooling on BEC. In addition, the present results will trigger a new route in developing ultra-high power gas laser with extremely narrow linewidth, or envisioning optical micro-mirrors and micro-cavities made with gas-phase materials. Among the interesting phenomena for investigations in the immediate future include molecular cooling via the observed resolved motional sidebands and engineering molecular non-classical state, and entangled photons to mention a few. Finally, the large induced acceleration in this laser-gas interaction could be explored as a new tool for micro- and nanoparticle accelerators.

**Acknowledgments and Author contributions**

FB incepted and developed the theory, designed the experiment and directed the research. MA fabricated the fibre, carried out the experiment and the data processing, AH contributed to the development of the theory and carried out the numerical simulations. FG and BD contributed to the data processing. All authors participated in discussions.

The authors thank A. Barthelemy for stimulating discussion and acknowledge support from “Agence Nationale de Recherche (ANR)” (grant PhotoSynth). MA is funded by King Saud University. AH acknowledges support from DFG, project HU1593/2-1.

# Methods

This is a supplemental document accompanying the manuscript “**Deeply-trapped molecules in self-nanostructured gas-phase material**” authored by Alharbi *et al*. Here, we provide the appendices and the associated figures or videos mentioned in the main manuscript.

**Appendix A: Experimental set-up for optical and RF spectral measurements.**

The 20 m long fibre used here is a home-made photonic bandgap (PBG) guiding HC-PCF and fabricated using the stack and draw technique. The fibre is filled with molecular hydrogen at a controllable pressure, by placing the two fibre-ends in gas cells. The gas pressure is kept uniform along the whole length of fibre by monitoring it with pressure gages placed at both cells. The gas cells are equipped with AR coated windows at both sides to avoid laser back reflection. The fibre has a core radius of $r$ = 3.2 μm (see top of Fig. 1B), and guides from 1000 nm to 1200 nm (Fig. 1C, red curve in the main manuscript), with a loss of 70 dB/km.

The hydrogen-filled fibre is pumped with a randomly polarized 1061 nm wavelength Yb-fibre CW laser that could emit up to a maximum of 100 W of optical power, with a linewidth of only ~400 kHz.

The optical spectrum from both fibre-ends is monitored using an optical spectrum analyser to record the FS and BS spectra.

Furthermore, the experimental set-up also comprises a portion with a self-heterodyne interferometer to measure the linewidth of both forward and backward propagating beam spectral components. The self-heterodyne system consists of a delay arm made of a 6 km long optical single-mode fibre (SMF), and a short modulation arm comprising an acousto-optic modulator (AOM) operating at ~211 MHz. The beat signal between the delayed optical beam and the AOM-frequency down-shifted signal is detected using a fast photo-detector (~1 GHz bandwidth) and recorded using a RF spectrum analyzer. The RF spectrum analyser has a resolution of 10 KHz when the span bandwidth exceeds 10 MHz. The linewidth traces of the pump, FS and BS are then recorded for different pump laser powers and gas pressures.

**Appendix B: Pump linewidth spectrum**

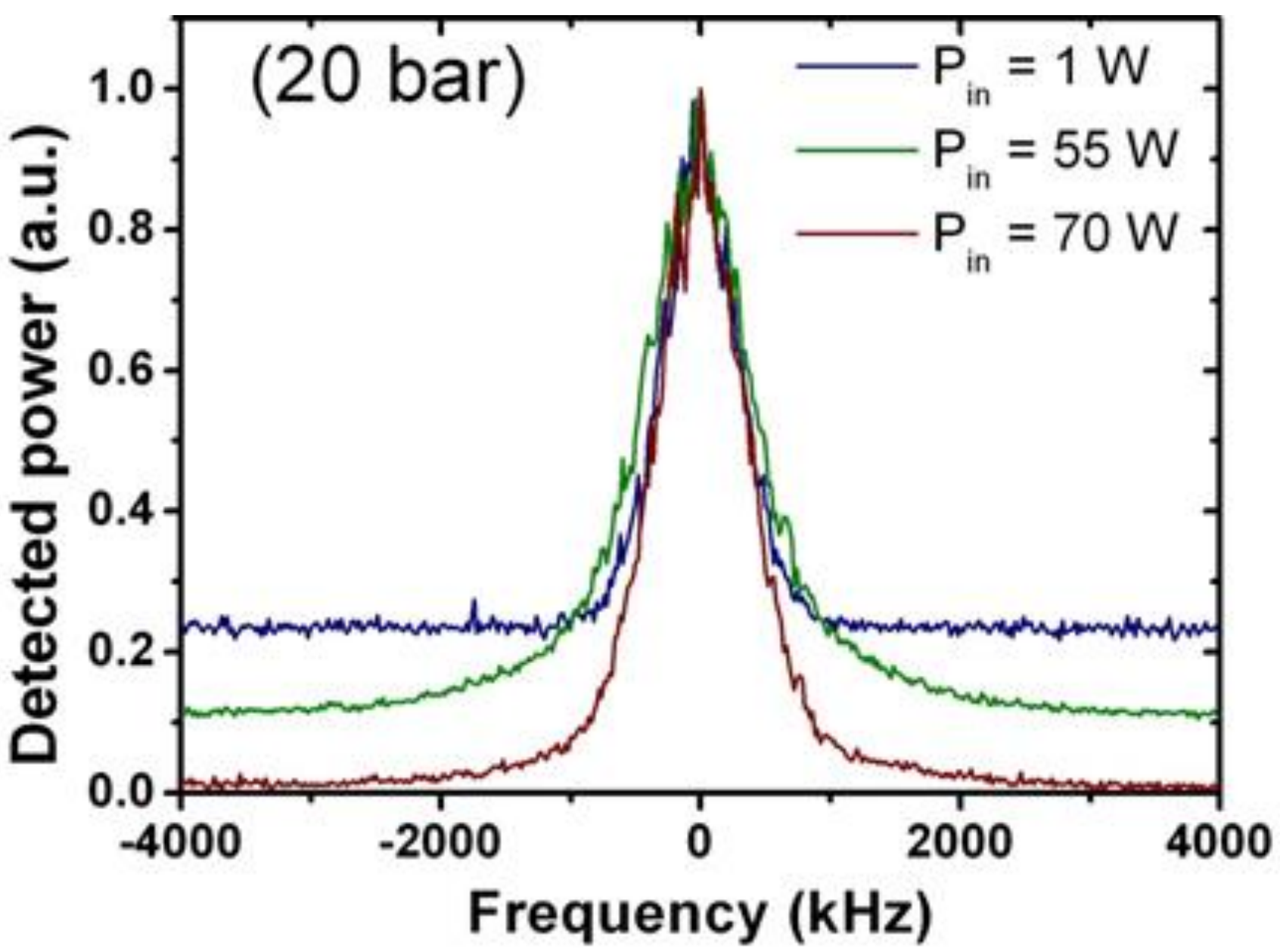

**Fig. M1** Pump linewidth for different input power at a pressure of 20 bar and fibre length of 20 m

The linewidth of the transmitted residual pump has been monitored for different input pump powers and gas pressures. Fig. M1 shows that the pump linewidth remains unchanged with input power increase.

**Appendix C: The theoretical model**

We consider the propagation of continuous-wave pump and first order Stokes radiation in the fundamental mode of the kagome-lattice photonic crystal fibre, neglecting the excitation of higher-order Stoke and anti-Stokes lines, as well as the energy transfer to higher-order transverse modes of the fibre.

The propagation characteristics of Stokes and pump, such as the wavenumbers $\beta_S$, $\beta_p$ as well as losses $\alpha_s$, $\alpha_p$, were calculated using the JCMwave finite-element Maxwell solver with high precision, using the tabulated data of the fused silica refractive index and the transverse cross-section of the fibre determined by the microscopy.

To derive the propagation equations, we first consider the steady-state values of the density matrix $\rho$, which are given through the coherence $\rho_{12}$ and the population difference $D = \rho_{22} - \rho_{11}$, where *1* and *2* denote the ground and excited rotational states of the $H_2$ molecules. At a fixed position, the electric fields of Stokes and pump components are given by $E_S(t) = (1/2)E_S\mathrm{e}^{-\mathrm{i}\omega_S\mathrm{t}} + c.c.$ and $E_p(t) = (1/2)E_p\mathrm{e}^{-\mathrm{i}\omega_p\mathrm{t}} + c.c.$ The steady-state values in this case are:

$$D = -\frac{\Gamma_{12}\gamma_{12}{}^2}{\Gamma_{12}\gamma_{12}{}^2 + 4|\Omega_{12}|^2\gamma_{12}}$$

$$\rho_{12} = \frac{i\Omega_{12}D}{\gamma_{12} - i(\Omega_{11} - \Omega_{22})},$$

where the Rabi frequencies $\Omega_{11}$, $\Omega_{22}$ and $\Omega_{12}$ are defined by

$$\Omega_{11} = 0.5(a_p|E_P|^2 + a_s|E_S|^2),$$

$$\Omega_{22} = 0.5(b_p|E_P|^2 + b_s|E_S|^2),$$

$$\Omega_{12} = 0.5 d_s E_P{}^* E_S.$$

Here $a_s$, $b_s$, $a_p$, $b_p$, $d_s$ are constants related to dipole moments. The quantities $\Gamma_{12}$ and $\gamma_{12}$ are the population decay rate and the Raman gain linewidth, correspondingly.

The Stokes component in the considered case consists of the forward-propagating part $E_S{}^{(f)}$ and backward-propagating part $E_S{}^{(b)}$, with $E_S = E_S{}^{(f)} + E_S{}^{(b)}$, which satisfy two distinct propagation equations.

The values $\rho_{12}$ and $D$ allow calculating the polarizations $P_{SF}$, $P_{SB}$ and $P_P$ as

$$P_S{}^{(f)} = 2N\hbar(a_S\rho_{11} + b_S\rho_{22})E_S{}^{(f)} + i\frac{N\hbar^2 d_s^2 D}{\gamma_{12} - i(\Omega_{11} - \Omega_{22})}|\mathrm{E_P}|^2 E_S{}^{(f)},$$

$$P_S{}^{(b)} = 2N\hbar(a_S\rho_{11} + b_S\rho_{22})E_S{}^{(b)} + i\frac{N\hbar^2 d_s^2 D}{\gamma_{12} - i(\Omega_{11} - \Omega_{22})}|\mathrm{E_P}|^2 E_S{}^{(b)},$$

$$P_P = 2N\hbar(a_P\rho_{11} + b_P\rho_{22})E_P + i\frac{N\hbar^2 d_S^2 D}{\gamma_{12} - i(\Omega_{11} - \Omega_{22})}|\mathrm{E_S}|^2\mathrm{E_P},$$

where $N$ is the concentration of the molecules.

The propagation equations are then written as

$$\partial_z E_S{}^{(f)} = -\frac{\kappa\omega_S}{2c\varepsilon_0}P_S{}^{(f)} - \alpha_S E_S{}^{(f)},$$

$$\partial_z E_S{}^{(b)} = -\frac{\kappa\omega_S}{2c\varepsilon_0}P_S{}^{(b)} + \alpha_S E_S{}^{(b)},$$

$$\partial_z E_P = -\frac{\kappa\omega_S}{2c\varepsilon_0}P_P - \alpha_P E_P,$$

where $\kappa = \kappa(E_P, E_S{}^{(f)}, E_S{}^{(b)})$ is the parameter which accounts for the microscopic spatial distribution of the gain. In the propagation equations, we have ignored the change of the refractive index which arises from microscopic density modulation, as detailed below. The analytical expression for κ is cumbersome and is not given here.

The origin of the backward Stokes component is the reflection from the input and output fibre interfaces back into the fibre due to mismatch of the effective refractive index of the fibre mode and of the free space, imperfections at the fibre ends, grating formed by the modulation of particle density, fibre roughness, with reflection coefficient estimated by a total value of 1% in energy. The boundary conditions for the Stokes field therefore are

$$\left|E_S{}^{(f)}(0)\right| = \sqrt{r}\left|E_S{}^{(b)}(0)\right|,$$

$$\left|E_S{}^{(b)}(L)\right| = \sqrt{r}\left|E_S{}^{(f)}(L)\right|,$$

where $r$ is the energy reflection coefficient and $L$ is the fibre length. We note, that the model doesn't take into account the reflection from the molecular lattice index modulation. Though neglecting the reflection off the Stokes optical lattice will have an effect on the exact magnitude of FS and BS, it doesn't significantly impact the results reported here.

Above equations were solved self-consistently with the propagation equations, using the numerical shooting method to determine $E_S^{(f)}(0)$. No reflection of the pump field was considered, since the values of the pump field at the fibre output were typically quite low. The following input parameters were assumed: $\omega_S$ = 1.6838 fs$^{-1}$, $\omega_P$ = 1.7933 fs$^{-1}$, $a_p$ = 3.854x10$^{-7}$ m$^2$/s/V$^2$, $a_s$ = 3.849x10$^{-7}$ m$^2$/s/V$^2$, $b_p$ = 3.85831x 10$^{-7}$ m$^2$/s/V$^2$, $b_s$ = 3.8536x10$^{-7}$ m$^2$/s/V$^2$, $d_s$ = 3.8538x10$^{-7}$ m$^2$/s/V$^2$, $\gamma_{12}$ = 2π (1.14x10$^9$)Hz, $\Gamma_{12}$ = 2π(2 x10$^5$)Hz, $N$ = 4.86x10$^{26}$ 1/m$^3$ at the pressure of 20 bar, L = 20 m, and waveguide core radius is 3.2 μm.

In our model, we consider the frequency difference between the pump and scattered Stokes to be equal to the Raman transition frequency, and the field-free population is in the ground state (*i.e.* $D = -\rho_{11} = -1$ ).

**Appendix D: FS and BS linewidth spectrum over narrow spectral span**

Fig. 3B of the main manuscript summarizes the results of FS and BS spectral emission linewidth and structure over $\pm 2$ MHz span along with the transmitted optical spectrum. The results are extracted from the full data set shown in figures M2-M5 Each figure shows, for a given fixed gas pressure, the evolution with the pump input power of both the optical spectrum of the transmitted beam, and that of 6 MHz-span RF spectrum of the forward and backward Stokes lines. Here, the input power ranges from 9 W (*i.e.* coupled power of 5.4 W) to 42 W. Given the necessity of resetting the laser fibre-coupling for each power level due to the change of its beam size and divergence, the number of input power runs was limited to 3 or 4 power-values of 9 W, 16 W, 29 W and 42 W. The figure series show the same content but at a different pressure. Four pressure values of 20 bar, 30 bar, 40 bar and 50 bar were investigated.

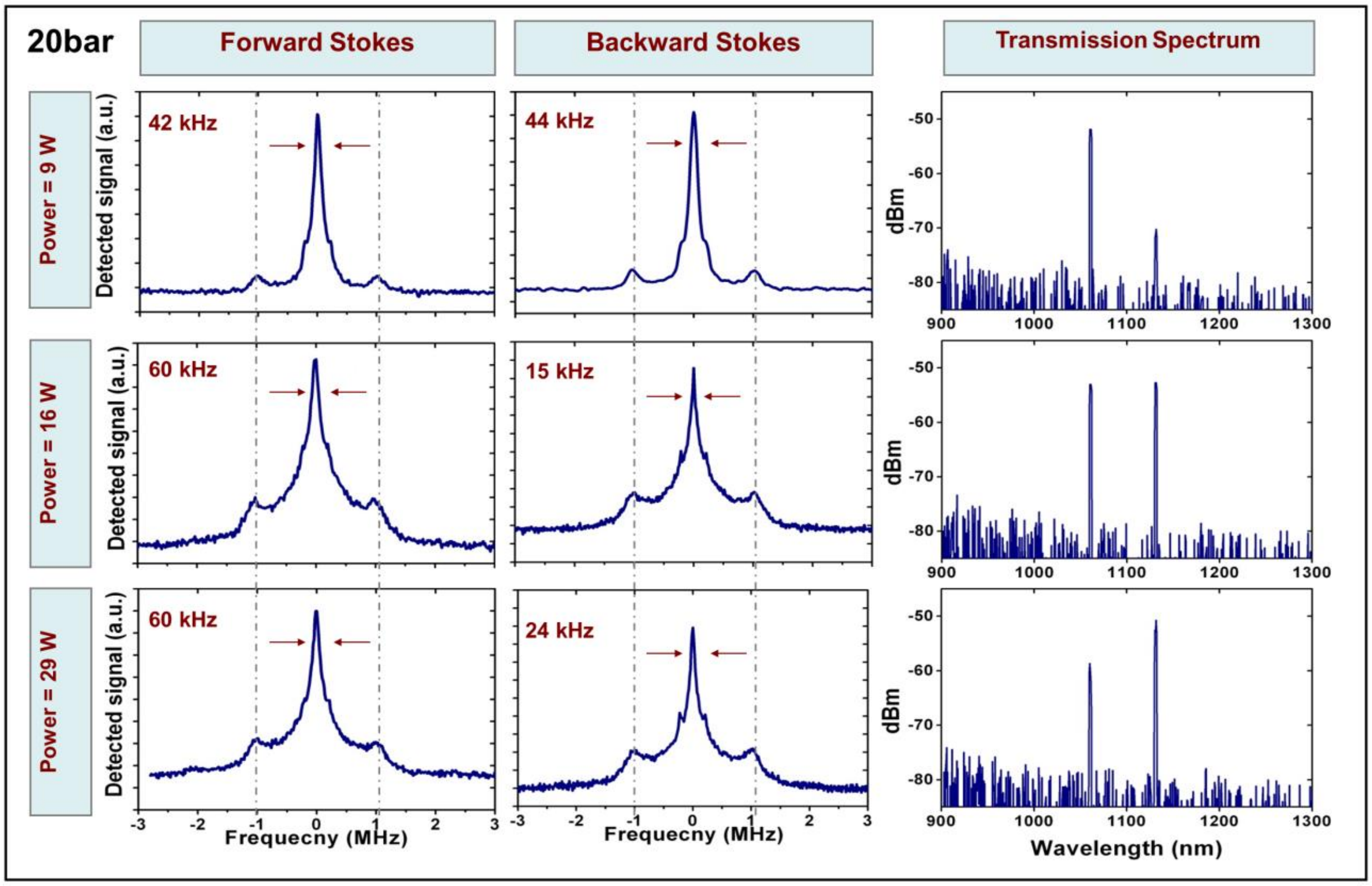

**Fig. M2** Linewidth traces of FS (first column) and BS (second column) and the optical spectral of the transmitted laser beam for different input power at a pressure of 20 bar and fibre length of 20 m. The measured linewidth is indicated in red.

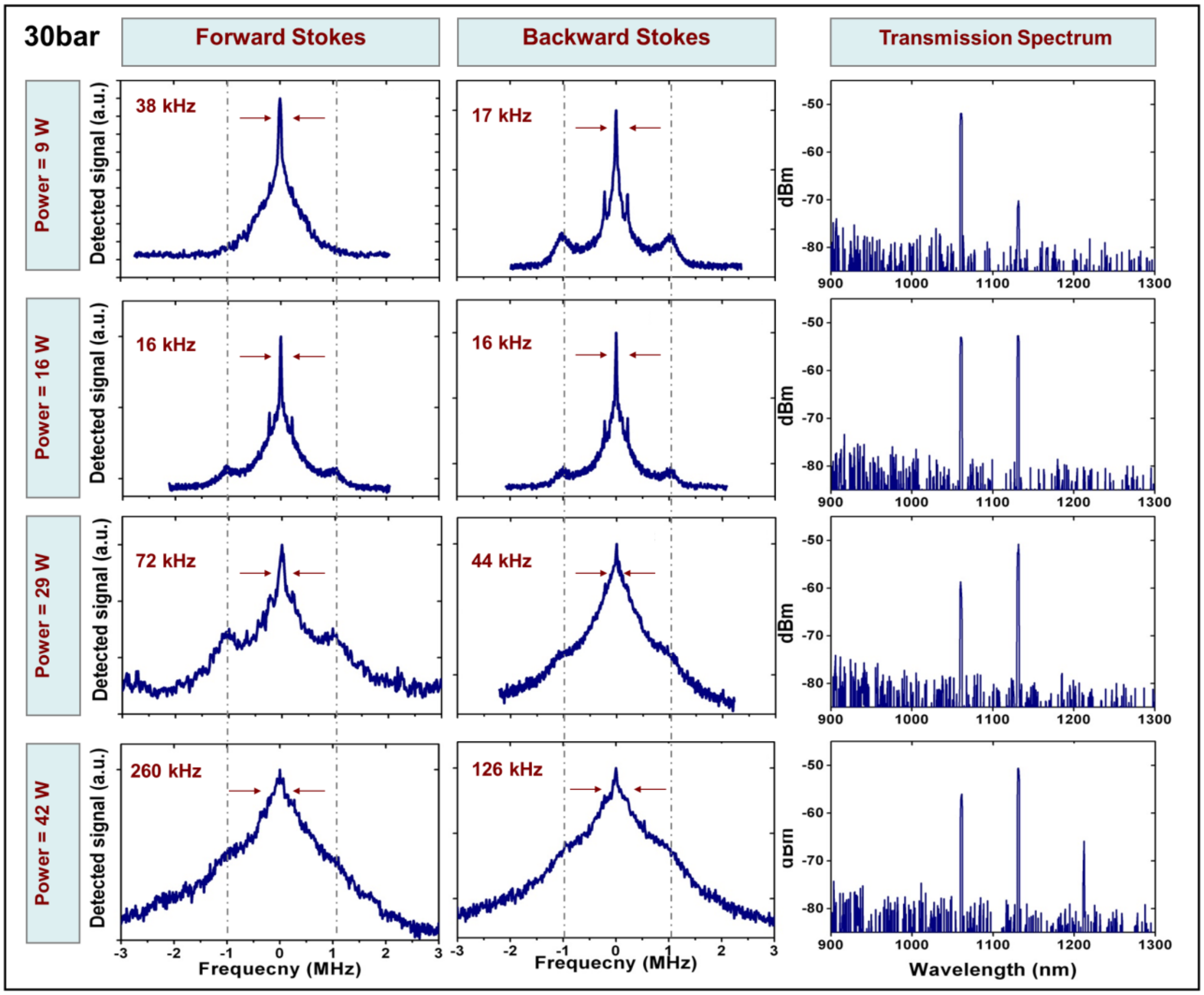

**Fig. M3** Same as in Fig. M2 but for a gas pressure of 30 bar.

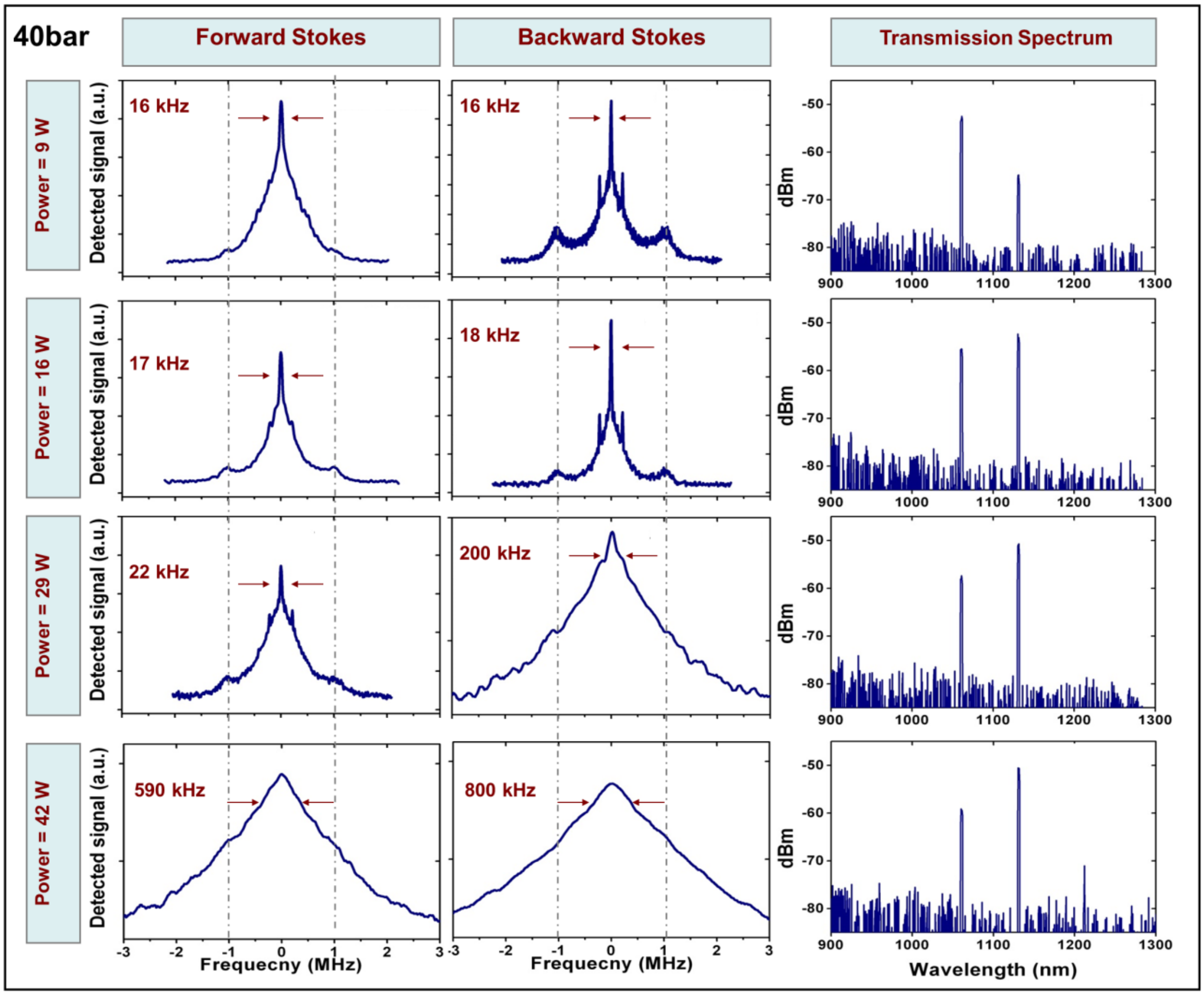

**Fig. M4** Same as in Fig. M2 but for a gas pressure of 40 bar.

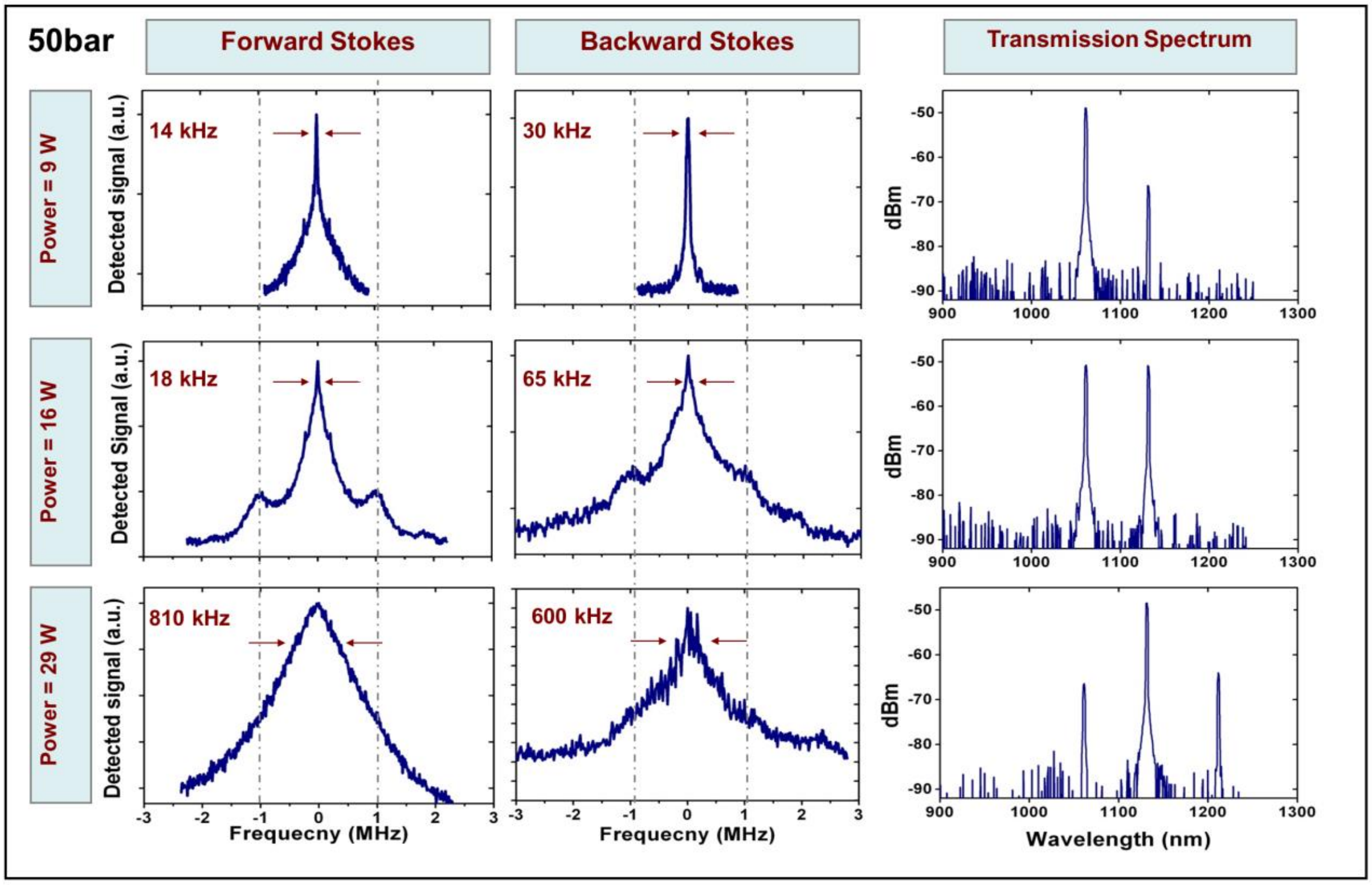

**Fig. M5** Same as in Fig. M2 but for a gas pressure of 50 bar.

## Appendix E: FS and BS linewidth spectrum over a broader spectral span

Here, we re-examine the linewidth traces and their evolution with input power and gas pressure, but over a larger RF frequency span of 150 MHz (resolution of 400 KHz). The recorded spectral traces are displayed in a similar fashion to those of the linewidth fine structure. Four figures (Fig. M6-M9) contain the evolution of the spectral traces with input power for a given fixed pressure. The pressure values are the same as in appendix D For the lower-range pressure (20 bar), the spectral traces of both FS and BS are dominated with a single narrow peak, which are accompanied by mainly two families of sidebands. The first sidebands are located in the range of Δ~12-15 MHz and their harmonic frequencies, and are identified as the two-photon Rabi sidebands (see manuscript). These sidebands vary little with input power and gas pressure due to the fact that the Raman active molecules are limited to those located in nano-lentils. The definition of the two-photon Rabi frequency is given in appendix G.

Furthermore, we observe that FS and BS exhibit a difference in the frequency of their sidebands in the range of sideband frequency of 1.2-2 MHz due to the overall motion of the molecular lattice. The second family of sidebands are located in the range of ~7-8 MHz, which is roughly half that of the Rabi sideband frequency. Moreover, higher order Rabi sidebands are also observed. We attribute this effect to four wave mixing (FWM) between Stokes central peak and the two sidebands and with a frequency given by $\omega_{2nd} = \omega_S + (\omega_S \pm \Omega_{Rs}) - (\omega_S \mp \Omega_{Rs})$=$(\omega_S \pm 2\Omega_{Rs})$. The 2nd order Rabi sideband signal will be determined by the nonlinear susceptibility at its frequency $\chi(\omega_{2nd})$ via $E_{2nd} = \chi(\omega_{2nd}) E_S(\omega_S) E_{Rs}(\omega_S \pm \Omega_{Rs}) E^*_{Rs}(\omega_S \mp \Omega_{Rs})$.

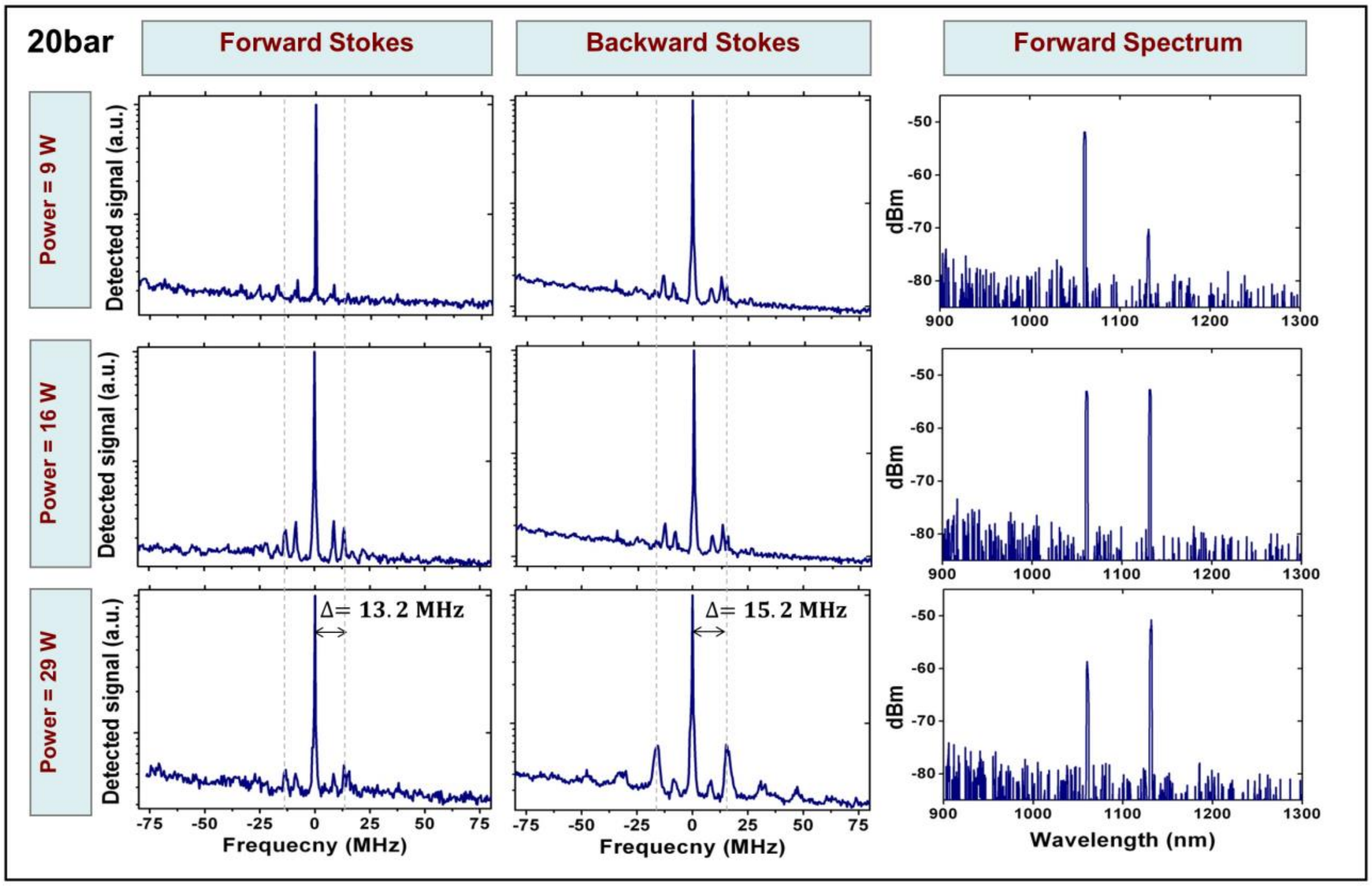

**Fig. M6.** 150 MHz span RF-spectra of Forward Stokes (first column) and Backward Stokes (second column) and the optical spectral of the transmitted laser beam (third column) for different input powers at a pressure of 20 bar and fibre length of 20 m. The measured two photon Rabi frequency is indicated for one representative case.

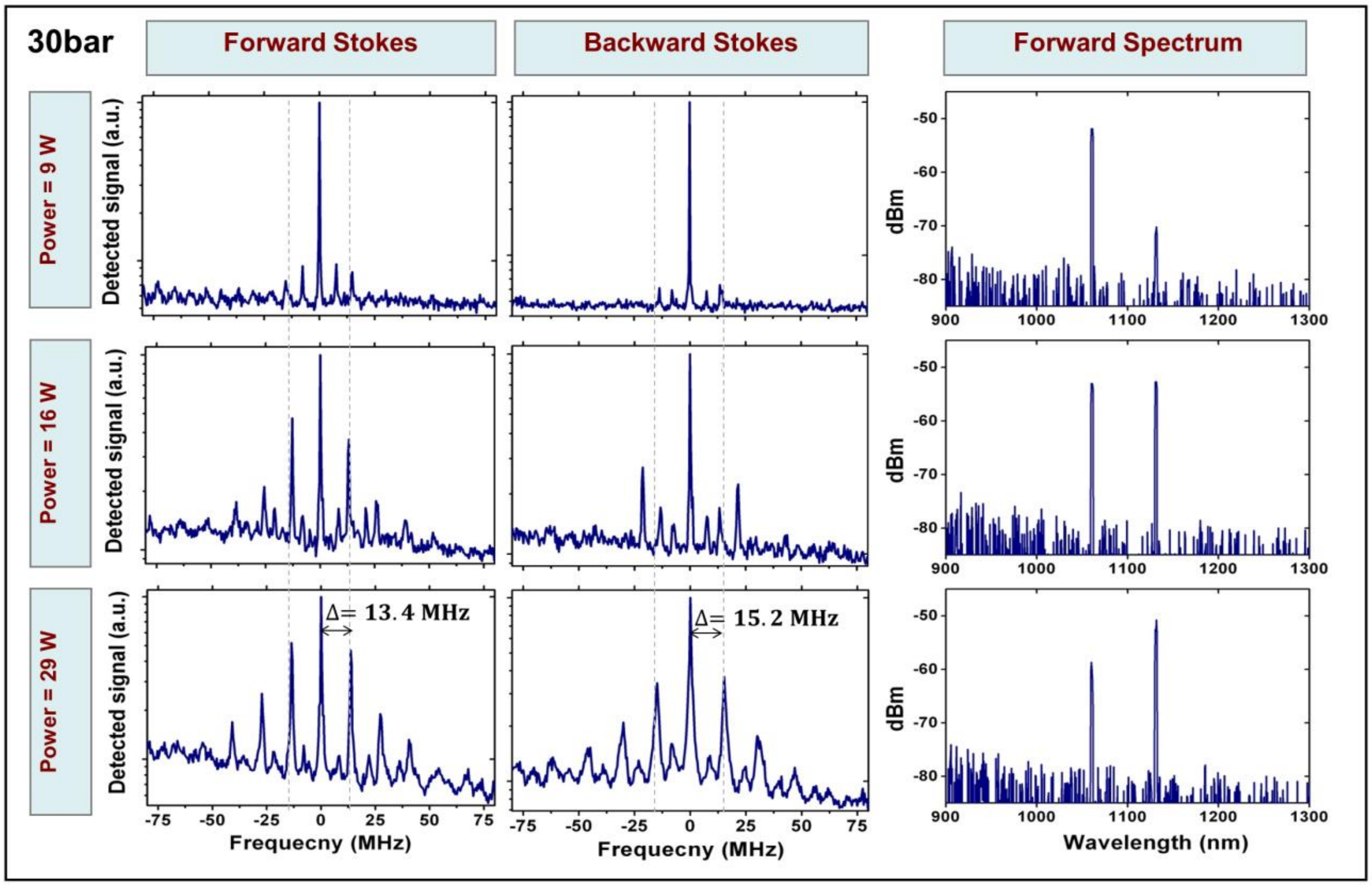

**Fig. M7.** Same as in Fig. M6 for gas pressure of 30 bar.

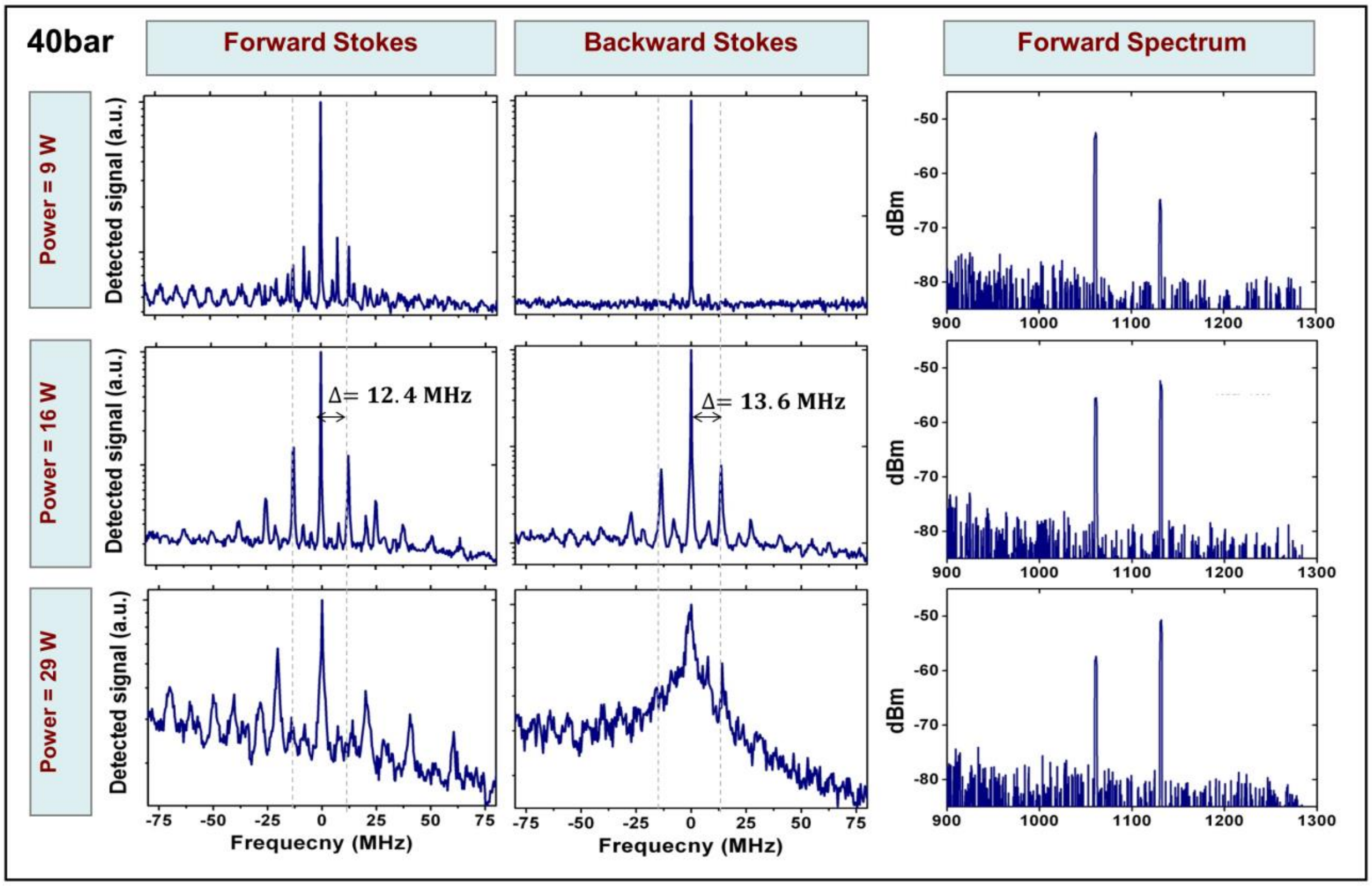

**Fig. M8.** Same as in Fig. M6 for gas pressure of 40 bar.

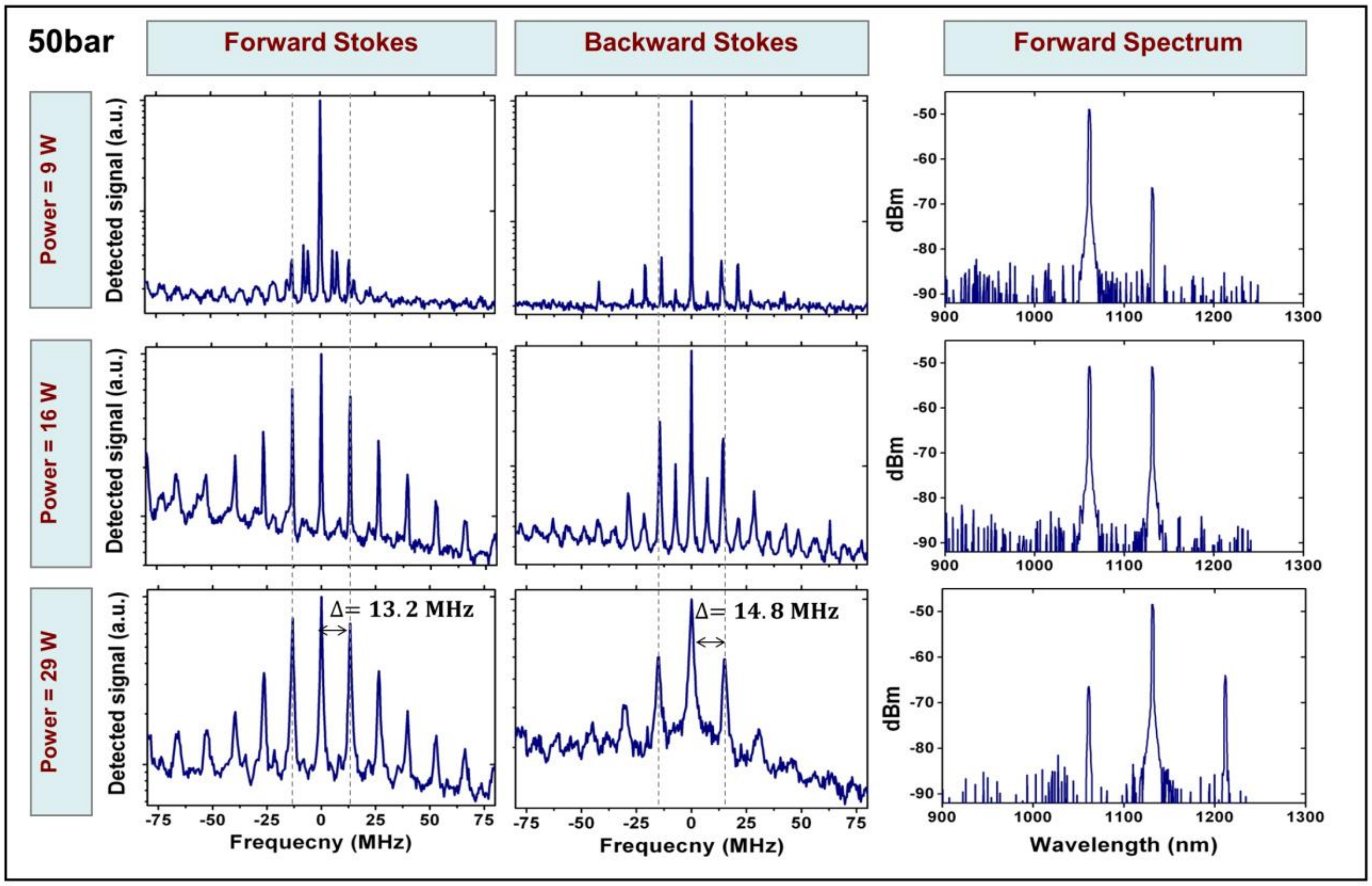

**Fig. M9.** Same as in Fig. M6 for gas pressure of 50 bar.

**Appendix F: Power scaling of ultra-narrow linewidth Stokes**

This appendix demonstrates the fibre power coupling handling and the extremely high quantum conversion to the first order Stokes.

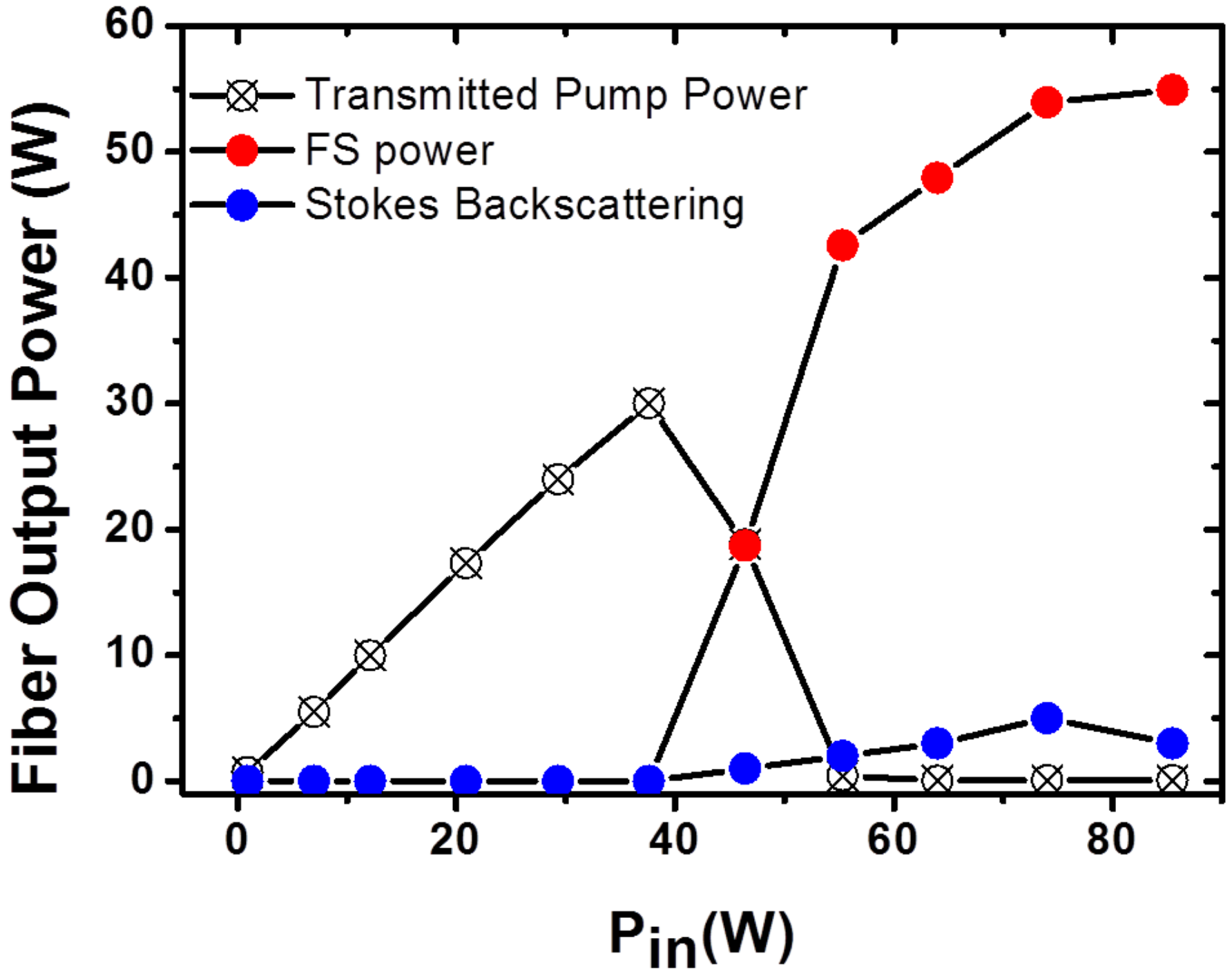

**Fig. M10**. Pump, FS and BS power evolution with input power for a fibre length of 7 m and a gas pressure of 20 bar.

Fig. M10 shows the evolution of FS and BS with input power for a different fibre length from the one considered in the main manuscript. Here the fibre length was set to 7 m. With this length we have demonstrated a coupling with input power as high as 85.5 W. At this input power, the FS power was found to be 55 W, and BS power 3 W. With the estimated fibre coupling efficiency of 75%, we find ~97% of quantum efficiency.

Furthermore, in the main manuscript, the ultra-narrow linewidth obtained with the 20 m long fibre were limited to input powers less than 30 W. Above this input power level the generation of the second order

Stokes strongly alters the SONS-GPM molecular lattice. This section is shown as a proof of concept that the lattice and hence the narrow linewidth can be obtained for higher input powers by simply shortening the fibre length and reducing the gas pressure. This will increase the input power onset at which the generation of the second-order Stokes occurs.

Fig. M11 shows the linewidth traces of FS and BS generated from a 7 meter long PBG HC-PCF. The fibre is similar to the one used so far. The linewidth measurements were taken with input power up to 70 W, and the gas pressure was set to 10 bar and 20 bar. We obtained with a pressure of 10 bar, a generated and transmitted Stokes with a power level in the range of 30-50 W and a linewidth of ~100 KHz.

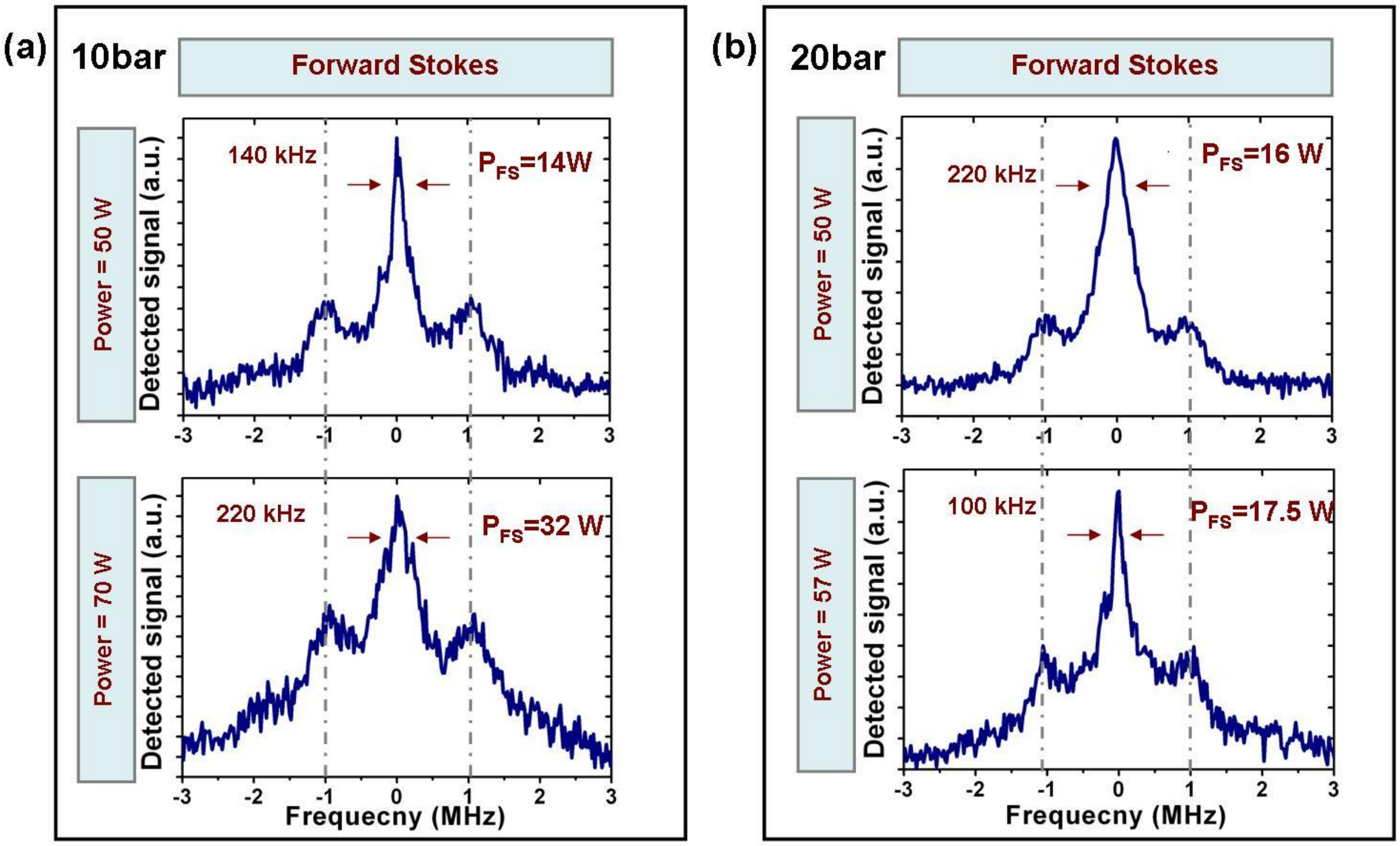

**Fig. M11.** (A) Linewidth for input power above 50 W for FS and BS using 7 m of PBG HC-PCF at gas pressure of (a) 10 bar and (b) 20 bar.

**Appendix G: Effective Rabi frequency definition**

The effective two-photon Rabi frequency $\Omega_{RS}$ acts only in the nano-lentils. Consequently, its expression deviates from $\Omega_{12}$, and its magnitude should be averaged over the wavelength, with the weight given by D. This gives the following expression:

$$\Omega_{RS} = \frac{\int_z^{z+\lambda} \Omega_{12}(z)D(z)dz}{\int_z^{z+\lambda} D(z)dz}.$$

**Appendix H: The phase diagram**

The phase diagram is defined as a map in the $(v, z)$ space of the value $E = \sqrt{v^2 + 2\langle U_{tot}(z)\rangle/m + C}$, with $\langle U_{tot}(z)\rangle$ given in the manuscript, *C* being a constant, and $E = -\sqrt{|v^2 + 2\langle U_{tot}(z)\rangle/m + C|}E$ for negative values of $v^2 + 2\langle U_{tot}(z)\rangle/m + C$. The adiabatically slow motion of a molecule corresponds to motion along the line of constant *E* in the *(*v*,z)* space. Moreover, since *E* has units of velocity, we have chosen $C = -2\langle U_{tot}(0)\rangle/m$ so that the value of *E* is equal to the velocity at *z = 0*. With such a definition, negative values of E mean localized motion.

**Appendix I: The influence of the particles redistribution on the dynamics**

As explained in the manuscript, the change of the gas density due to the modulated expectation value of the Hamiltonian leads to the significant, above fourfold, increase of the density in the lentils. This higher density will lead to two major effects: first, the population decay rate and the coherence decay rate are going to be modified. Second, the higher density will result in a higher collision rate of the molecules. Both of these effects, in turn, lead to the modification of the expectation value of the Hamiltonian: the former one directly as described by the formalism shown in the paper, the second one indirectly, by modifying the time molecules dwelling in any given position under the influence of the field. Therefore the distribution of the molecules over the position, velocities, and the quantum state described by the $\rho_{12}$ and *D* should be calculated self-consistently, including the above effects. This calculation is not included in the current simplified version of theory; however, even such a simplified version gives quantitative agreement to the experimental values.

**Appendix J: Movie showing the moving scattering nanoparticles inside the hydrogen filled HC-PCF**

Fig. M12 shows a snap-shot from a video showing a scattering dust that has been trapped in the fibre guided beam. This was achieved by imaging a fibre section that was set in a spiral form so as to capture as much length as possible within a single frame. The imaging was recorded using a CCD camera in front of which is mounted an IR viewer. The speed of the scatterers was deduced be recording their location within the frame in function of time, deduced from the frame number and the frame rate. The full video is available in the online supplementary materials.

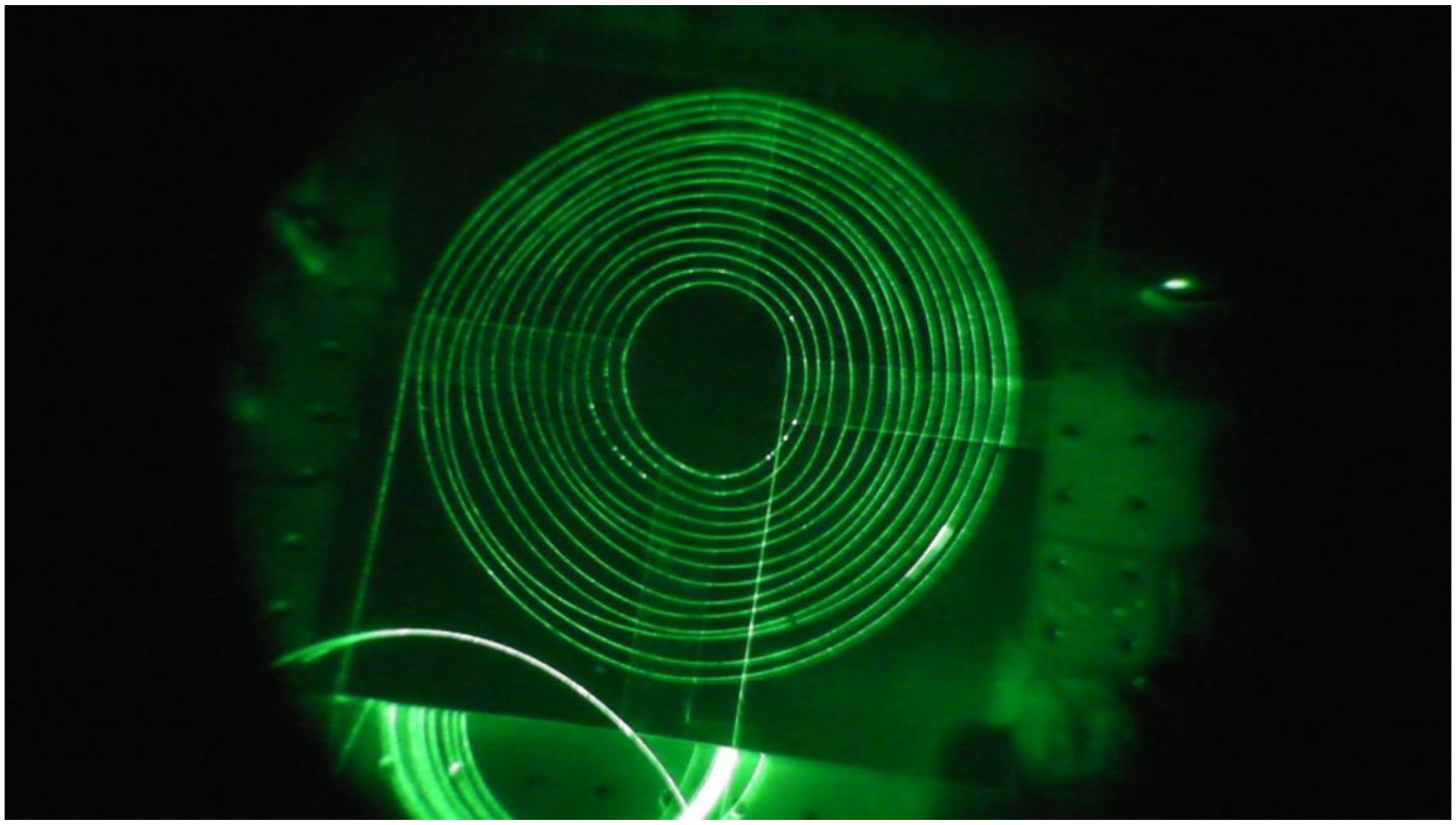

**Fig. M12.** A snapshot of a recorded video for moving scattering particles at input power of ~ 29 W (video available online).